\documentclass[usenatbib]{aastex631}

\begin{document}

\title[QSO PG1004+130]{PG1004+130: Hybrid Morphology Source or a Restarted FRII? A uGMRT Polarimetric Investigation}

\author{Salmoli Ghosh}
\altaffiliation{salmoli.ghosh@gmail.com}
\affiliation{National Centre for Radio Astrophysics (NCRA) - Tata Institute of Fundamental Research (TIFR)\\ S. P.   Pune University Campus, Ganeshkhind, Pune 411007, India}

\author{P. Kharb}
\affiliation{National Centre for Radio Astrophysics (NCRA) - Tata Institute of Fundamental Research (TIFR)\\ S. P.  Pune University Campus, Ganeshkhind, Pune 411007, India}

\author{J. Baghel}
\affiliation{National Centre for Radio Astrophysics (NCRA) - Tata Institute of Fundamental Research (TIFR)\\ S. P.  Pune University Campus, Ganeshkhind, Pune 411007, India}

\author{Silpa S.}
\affiliation{National Centre for Radio Astrophysics (NCRA) - Tata Institute of Fundamental Research (TIFR)\\ S. P.  Pune University Campus, Ganeshkhind, Pune 411007, India}

\begin{abstract}
We present here the polarization image of the hybrid morphology (HYMOR) and broad-absorption line (BAL) quasar PG1004+130 at 694~MHz obtained with the upgraded Giant Metrewave Radio Telescope (uGMRT). We detect linear polarization in this source's core, jets, and lobes. The visible discontinuity in total intensity between the inner jets and the kpc-scale lobes suggests that the source is restarted. The inferred poloidal magnetic (B-) field structure in the inner jet is consistent with that observed in Fanaroff-Riley (FR) type II sources, as are the B-fields aligned with the lobe edges. Moreover, archival Chandra and XMM-Newton data indicate that PG1004+130 displays several FRII-jet-like properties in X-rays. We conclude that PG1004+130 is a restarted quasar, with both episodes of activity being FRII type. The spectral index images show the presence of an inverted spectrum core ($\alpha=+0.30\pm0.01$), a steep spectrum inner jet ($\alpha=-0.62\pm0.06$) surrounded by much steeper lobe emission ($\alpha\approx-1.2\pm0.1$), consistent with the suggestion that the lobes are from a previous activity episode. 
The spectral age difference between the two activity episodes is likely to be small ($<1.2 \times 10^7$ years), in comparison to the lobe ages ($\sim 3.3\times 10^7$ years). The inferred B-fields in the lobes are suggestive of turbulence and the mixing of plasma. This may account for the absence of X-ray cavities around this source, similar to what is observed in M87's radio halo region. The depolarization models reveal that thermal gas of mass $\sim (2.4\pm0.9)\times 10^9$ M$_\sun$ is mixed with the non-thermal plasma in the lobes of PG1004+130.
\end{abstract}


\section{Introduction}
Active galactic nuclei (AGNs) are understood to be mass-accreting supermassive black holes (SMBHs; $\sim10^6 - 10^9$ M$_{\sun}$) in the centre of galaxies \citep{Rees1984}. At the interface of the SMBH and accretion disc, relativistic jets of synchrotron-emitting plasma are launched, which can exhibit differing powers and morphologies at radio frequencies. These radio jets, extending from parsec to Mpc scales \citep{Blandford2019}, remain collimated by extracting energy and angular momentum either from the central SMBH or the accretion disc \citep{Blandford1977,Blandford1982}. \citet{Fanaroff1974} found that AGN with large kpc-scale jets exhibited a dichotomy: the lower radio luminosity ($L_{\rm{178\,MHz}} < 2\times 10^{25}$ W~Hz$^{-1}$~sr$^{-1}$) sources had bright cores and plume-like radio lobes and were classified as Fanaroff-Riley type I (FRI) while higher radio luminosity ($L_{\rm{178\,MHz}} > 2\times 10^{25}$ W~Hz$^{-1}$~sr$^{-1}$) sources showed collimated jets with terminal hotspots and were the FR type II (FRII) sources. However, based on different samples, several authors have suggested that the luminosity-based division is not that strong \citep{Kharb2010,Mingo2019}. Instead, properties such as the jet shape, flow pattern, and jet composition (i.e., electron/proton content) on kpc-scales may mark the true differences between the sub-classes \citep{Laing2002,Croston2018}. In radio polarimetric observations, powerful FRII sources tend to have the inferred magnetic (B-) fields parallel to the jet flow axis, whereas FRI sources typically display jets with the B-fields predominantly perpendicular to the jet flow axis or B-fields perpendicular to the jet axis at the centre of the jet, and parallel near one or both of its edges \citep{Willis1981,Bridle1982}. 
 
A small fraction of radio galaxies show an FRI jet on one side of the radio core, and an FRII jet on the other, and are referred to as hybrid morphology (HYMOR) sources \citep{Gopal-Krishna2000, Gawronski2006,Kapinska2017}. The formation mechanisms of HYMORs have been widely debated. Their very existence could challenge the role of the central engine in producing FRI or FRII-type jets, as both types of jets are launched from the same engine in these sources. It, therefore, directs the focus toward the role of the environment around the sources. The surrounding media may be asymmetrically distributed, resulting in greater jet-medium interaction and thereby de-collimation in one of the jets (on the FRI side) \citep{Gopal-Krishna2000,Stroe2022}. HYMORs can also be observed as such due to projection effects, restarted AGN activity, Doppler boosting in jets, or other factors \citep{Harwood2020,Sobha2022,Gopal-Krishna2023}. This paper presents the results from radio polarization observations of a `HYMOR' source, PG1004+130, using the uGMRT.

\subsection{PG1004+130}\label{PG1004+130}
PG1004+130 (also known as 4C+13.41) has been classified as a rare example of a HYMOR \citep{Gopal-Krishna2000,Gawronski2006,Miller2006,Scott2015}, that is also a broad-absorption line (BAL) quasi-stellar object (QSO).  BALs are exhibited by several QSOs in their ultraviolet (UV) spectrum \citep{Hall2013}. These are usually observed in the high-ionisation lines like CIV, SiIV, NV and suggest outflows with velocities upto 30,000~km~s$^{-1}$ \citep{Reichard2004, Goodrich1995}. BALs are observed when our lines of sight are directly along these wind-like outflows \citep{Elvis2000}, and are typically observed in radio-quiet (RQ) AGNs \citep{Goodrich1995, Stocke1992}. 
PG1004+130 is the first radio-loud \citep[RL; where the radio loudness parameter, $R\equiv {S_{5\,\mathrm{GHz}}/S_{B- \mathrm{band}}}\sim210$, $S_{\nu}$ being the flux density at frequency $\nu$;][]{Kellermann1989} BAL QSO discovered at a redshift (z) of 0.24 \citep{Wills1999}. 
The kpc-scale radio structure comprises a hotspot-like high surface brightness region towards the northern lobe edge consistent with the classical `edge-brightened' definition of an FRII source, while the southern lobe is more plume-like with no signature of a hotspot, similar to an FRI source \citep[see][]{Fomalont1981,Kunert-Bajraszewska2013,Baghel2023}.
At 178~MHz, the source has a flux density of $\sim5$~Jy \citep{Parkescatalog1990}, or a luminosity of $\sim 1.8\times 10^{26}$ W Hz$^{-1}$ sr$^{-1}$ given its luminosity distance of 1244.7 Mpc\footnote{NASA NED}, placing it into the FRII category. 

PG1004+130 is hosted by a giant elliptical galaxy \citep{O'Brien1998} with an SMBH of mass of $2.7\times10^{9}$~M$_{\sun}$ \citep{Shangguan2018}. X-ray observations of PG1004+130 with the \textit{Chandra} and the \textit{XMM-Newton} telescopes reveal an X-ray jet and asymmetric diffuse X-ray emission coincident with the radio lobes \citep{Miller2006,Scott2015}. No X-ray cavities \citep[e.g.,][]{Vantyghem2014} are detected till now near the kpc-scale radio emission of PG1004+130 \citep[e.g.,][]{Miller2006,Scott2015}. Additionally, there is no X-ray emission from the hotspot. Hard X-ray observations with the Nuclear Spectroscopic Telescope Array (\textit{NuSTAR}) by \citet{Luo2013} recommend PG1004+130 to be significantly X-ray weak at hard X-rays (20-30 keV); they suggest that PG1004+130 could have Compton-thick shielding gas in its nuclear regions. 

In this paper, we study the low-frequency observations of this HYMOR in light of its known multi-waveband properties to infer the nature of hybrid morphology sources. The cosmology used throughout is H$_0$ = 67.8~km~s$^{-1}$~Mpc$^{-1}$, $\Omega_\mathrm{mat}$ = 0.308, $\Omega_\mathrm{vac}$ = 0.692, and $\Omega_k$ = 0 (flat universe). At the distance of PG1004+130, 1$\arcsec$ corresponds to a projected linear size of $\sim$ 3.9 kpc. The spectral index, $\alpha$, is defined such that $S_\nu \propto \nu^\alpha$.

\section{Radio Data: Observations \& Reduction} \label{section:data analysis}
We acquired the uGMRT observations in Band-4 (694 MHz) on 2019 December 30 (Project ID: 37\_042, PI: Silpa S.) with a resolution of $\sim4.3\arcsec$. The on-source time was 49 minutes. 3C286 has been used as the polarization angle and flux calibrator, 3C48 as the `unpolarized' polarization calibrator at 694 MHz, and 1120+143 as the phase calibrator. While the `unpolarized' calibrator OQ208 was also observed, it could not be used as a polarization calibrator due to its low total flux density at the uGMRT Band-4 frequencies. The primary data reduction and imaging were carried out using the \textsc{Common Astronomical Software Application} \citep[CASA;][]{CASA2022}. We also used the 
\textsc{Astronomical Image Processing System} \citep[\textsc{AIPS};][]{vanMoorsel1996}, along with \textsc{casa}, for carrying out several of the post-imaging analysis tasks, which are explicitly mentioned below.

For setting the flux density of 3C286 (22.37 Jy at 0.559 GHz) and 3C48 (31.82 Jy at 0.559 GHz), the \citet{Perley2017} scale was adopted. The basic rounds of calibrations, such as the phase, delay, bandpass and complex gain calibrations, were done following the standard procedures. The polarization calibration was done in three steps: cross-hand delay calibration, `leakage' (D-term) calibration and the polarization angle calibration \citep[for details, see][]{Silpa2021,Baghel2022}. We had acquired only two scans on 3C286 in these observations, which were insufficient for solving the frequency-dependent D-terms. Therefore, we used the `unpolarized' calibrator 3C48 to solve for the polarization leakages. The average instrumental leakage for our data was around $15\%$ \citep[see also][]{Silpa2021,Baghel2022}. 

After the application of the calibration tables to the visibility dataset and some additional flagging on the target field, the calibrated target source visibilities were extracted using the task \textsc{split} in \textsc{casa} with the averaged spectral channels (7 at a time), to decrease the data volume without causing bandwidth smearing. The dirty Stokes I, Q, U images of the target were obtained with the \textsc{tclean} task in \textsc{casa} using the \textsc{multi-term multi-frequency synthesis} \citep[mtmfs;][]{Rau2011} deconvolver. Subsequently, four phase-only and four amplitude+phase self-calibration cycles were run on the Stokes I image. The final Stokes Q and U images were made from the self-calibrated visibilities. The r.m.s. noise in the final Stokes I, Q, U images are (120, 87, 66) $\mu$Jy~beam$^{-1}$, respectively. The above steps are included in a Python script as an \href{https://github.com/jbaghel/Improved-uGMRT-polarization-pipeline}{improved uGMRT polarization pipeline.}

The flux densities, obtained from the Stokes I image, for the different regions of PG1004+130 at 694~MHz have been listed in Table \ref{tab:2}. The flux density errors have been calculated using the formula $\sqrt{\left(\sigma\sqrt{N_b}\right)^2+(\sigma_p S)^2}$ \citep{Kale2019}, where $\sigma$ is the r.m.s. noise, $N_b$ is the number of beams in the region, $S$ is the flux density of that region and $\sigma_p$ is the percentage error in flux density. For compact regions, like the core, the peak intensity has been noted to avoid contamination from the surrounding extended emission.

\subsection{Polarization Intensity Image}
The Stokes I, Q, and U images, obtained as described above, give us the linear polarization information of radiation from the source. We obtain the linear polarization intensity (I$_P$) as $\sqrt{\mathrm{Q}^2+\mathrm{U}^2}$ and the linear polarization fraction (FP) as $\sqrt{\mathrm{Q}^2+\mathrm{U}^2}/\mathrm{I}$. The polarization angle ($\chi$) values were obtained as $0.5\rm{\tan^{-1}}{\left(\mathrm{U}/\mathrm{Q}\right)}$. In \textsc{aips}, the I$_P$ image was made with the task \textsc{comb} and opcode \textsc{polc} (considering Ricean bias correction) on the Q and U maps, clipping below the signal-to-noise ratio (SNR) of 3. The $\chi$ image was made with \textsc{comb} using opcode \textsc{pola} on the Q and U maps, clipping above a noise ($\chi$ error) of 10$\degr$. To get the FP image, we used \textsc{comb} with opcode \textsc{div} on the I$_P$ and total intensity maps, clipping above a noise (FP error) of 10\%.

\subsection{Spectral Index Images}\label{secspectral}
The two frequency spectral index images were created using the Stokes I images at the two frequencies made with the identical image size, cell-size and beam-size in \textsc{tclean} in \textsc{casa} or \textsc{imagr} in \textsc{aips}, and subsequently using the task \textsc{comb} in \textsc{aips} with opcode \textsc{spix} and clipping below 3 times the r.m.s. noise. Two spectral index images along with error images were made by combining data at 694~MHz from the uGMRT and archival 1.5~GHz data from the historical Karl G. Jansky Very Large Array (VLA) B-array (resolution $\sim4\arcsec$; legacy ID: AK219, 1989 March 18), and 6~GHz VLA B-array data \citep[][]{Baghel2023} with archival 1.5 GHz data from the VLA A-array (resolution $\sim 1.4\arcsec$; legacy ID: WARD, 1982 March 4). 
The spectral index images were masked with the corresponding error images using the task \textsc{blank} in \textsc{aips}.

\subsection{Rotation Measure Image}
The rotation measure (RM) image was created in \textsc{casa} with the help of the task \textsc{rmfit}, with the QU image cubes at the different frequencies as the inputs. Images at 3 frequencies were taken to avoid the n$\pi$ ambiguity in the RM estimates. The VLA 6~GHz data with a total bandwidth of 2~GHz were split into two sub-bands (central frequencies: 4.9~GHz, 5.9~GHz), each having 512 channels (8 spectral windows) and then imaged. Each of the split images show significant amount of polarization. The \textsc{rmfit} task produced a pixel-to-pixel RM map as defined by the difference in the EVPAs at the different wavelengths divided by the difference in the squared wavelengths using the $\chi$ images at 694~MHz, 4.9~GHz, 5.9~GHz. In order to obtain the RM value for the core, we could use the two uGMRT sub-band (centred at 632 MHz, 757 MHz) and one of the VLA sub-band (5.9~GHz) QU image cubes. Contrary to the core, polarization was not detected in the lobes for the uGMRT\footnote{Total bandwidth of uGMRT data was only 250~MHz.} sub-band images and therefore could not be used for RM estimation. The RM images were then blanked by the RM error images with the task \textsc{blank} with a suitable flux cutoff at a given frequency. 

All these images (total intensity, polarization intensity, spectral index, and rotation measure) were then examined using \textsc{casa}. In \textsc{casa}, the region of interest was chosen with a polygon selector tool, and then the mean values were noted by checking the statistical analysis. This gave us an average value of the quantities corresponding to the images in the region of choice. 

\section{Results} \label{section:results}
\subsection{The Radio Structure}
\begin{figure*}
    \centering
    \includegraphics[trim={2.6cm 0 1.2cm 0},clip,width=18cm]{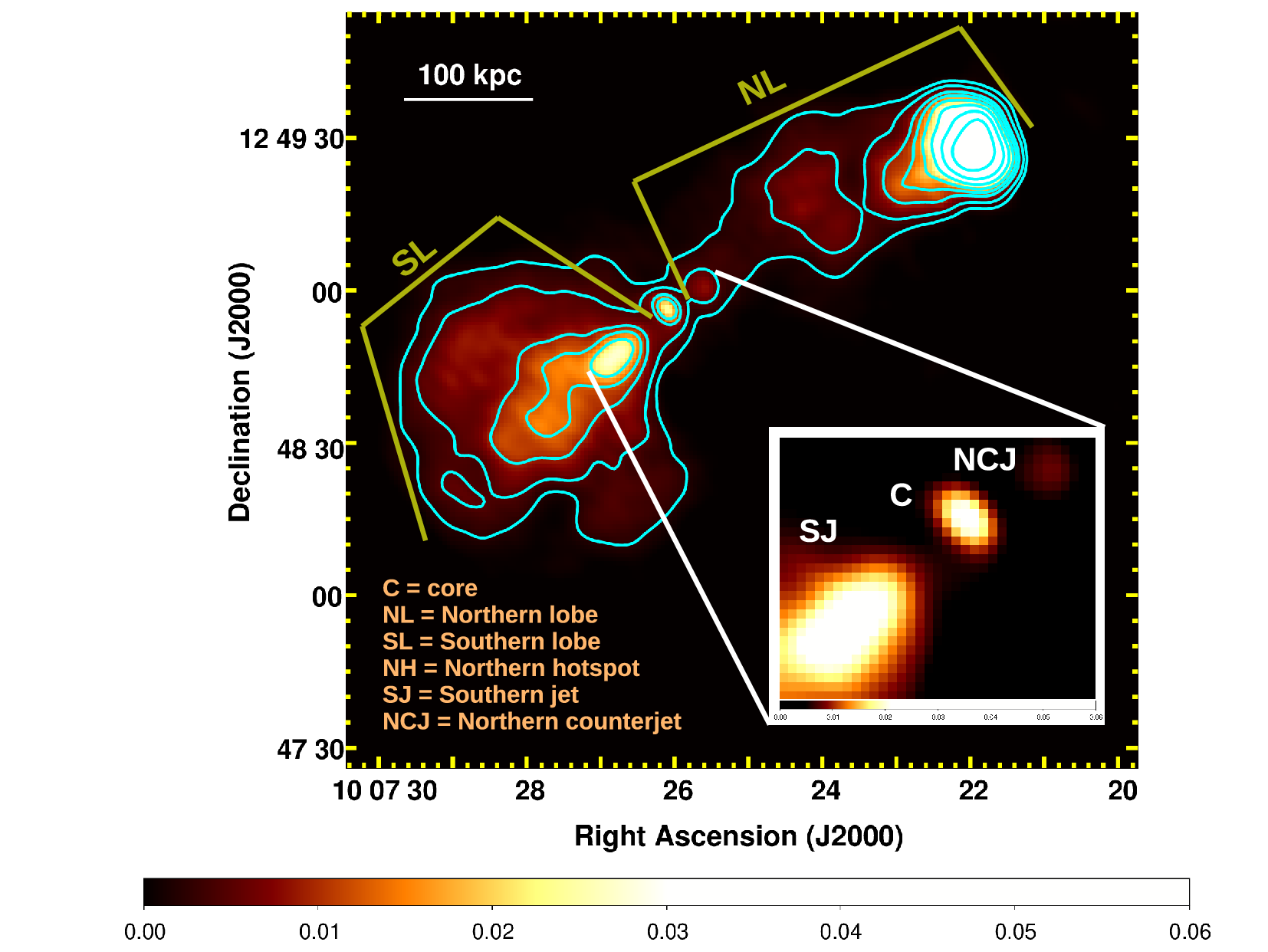}
    \caption{uGMRT 694 MHz total intensity image of PG1004+130, showing different radio features as listed in the image. The contour levels are at (20, 40, 80, 113.14, 160, 226.27, 320, 452.55) $\times$ 120 $\mu$Jy~beam$^{-1}$. The inset shows a zoomed-in version of the jet-core-counterjet structure. The surface brightness in colour ranges from 0 to 60 mJy~beam$^{-1}$.} 
    \label{fig:total intensity}
\end{figure*}

The radio emission from PG1004+130 is much more extensive ($\sim 137\arcsec$ or $\sim539$~kpc) than the giant elliptical host galaxy, which has an extent of $\sim$ 10$\arcsec$ ($\sim$ 39~kpc), as measured using the task \textsc{tvdist} in \textsc{aips}.  
The uGMRT 694 MHz image shows clear sub-stuctures. We have classified the radio structure into various sub-components as in Figure~\ref{fig:total intensity}: the Northern hotspot (NH), the Northern lobe (NL), the Southern lobe (SL), the Southern jet (SJ), the core (C) and the Northern counterjet (NCJ). We have followed this nomenclature in the rest of the paper. We note here that we do not detect actual counterjet emission from NCJ and only see the hotspot associated with the same. This is because the counterjet emission is likely to be Doppler-de-boosted as the jet is receding away from our line of sight. Indeed, VLBI images clearly show that the approaching jet is towards the South in PG1004+130 \citep{Wang2023}. The total flux density at 694~MHz for PG1004+130 is 2.2~Jy. The flux densities and the peak brightnesses for different regions are tabulated in Table~\ref{tab:2}. 

\subsection{Polarization and B-fields}
We present the RM-corrected polarization image of PG1004+130 in Figure~\ref{fig:pol}. The polarization information about PG1004+130 is provided in Table~\ref{tab:1}. All the values in Table~\ref{tab:1} represent the mean values over the mentioned regions.
The $\chi$ values are tabulated after a global RM-correction to the EVPAs. The EVPAs were rotated by $-$100 degrees corresponding to an RM of $-9.5$~rad~m$^{-2}$, so that it matched the EVPAs observed in the VLA 6~GHz image of \citet{Baghel2023}, assuming the latter to not be significantly affected by Faraday rotation on account of the higher frequency. In doing so we find that the EVPAs in the Southern jet match between the GMRT and VLA images as well, giving us confidence about the validity of our approach and the resultant RM value. As further validation, the resultant RM value falls broadly around the global RM range reported in the literature for this source \citep[RM $=-16\pm5$~rad~m$^{-2}$;][]{Simard1981}.

We have not attempted to correct for any ionospheric Faraday rotation. This is primarily due to the lack of robust ionospheric models for the GMRT. Using Global Positioning System (GPS) measurements of the vertical total electron content (TEC) from stations in nearby cities and assuming the magnetic field to be weakly variable over the course of their observations, we note that \citet{Farnes2014} concluded that the Faraday rotation due to the ionosphere was $\le2$~rad~m$^{-2}$ for their GMRT observations.

The uGMRT image picks up extensive diffuse emission with complex polarization features compared to the VLA 6~GHz image. The degree of linear polarization ranges from $4-18$\% across the source. The Northern hotspot is highly polarized with a FP value of $\sim$ 15\%. The red ticks in Figure \ref{fig:pol} denote the EVPA vectors, with the length of the ticks being proportional to I$_P$ values. The inferred B-field structures for the extended and optically thin regions like the Southern jet or/and lobes are shown as insets in Figure \ref{fig:pol} by rotating the EVPA vectors by 90$\degr$ following \citep[Chapter 3,][Section 5.3.2]{Pacholczyk1970,Ioannis2015}. The EVPA vectors are perpendicular to the jet direction in the southern inner jet. Interestingly, there is no compact hotspot-like structure with EVPA vectors parallel to the jet direction at the end of the outer lobe. We discuss this further in Section~\ref{sec:discussion}.

\begin{table*}
    \centering
    \caption{Summary of polarization properties of PG1004+130}
    \label{tab:1}
    \tabcolsep=0.11cm
    \begin{tabular}{ccccc}
    \hline
    \hline
         Region&I$_P$&$\chi$&FP & RM\\
          &(mJy/beam)&(degrees)&(\%) & (rad~m$^{-2}$) \\
         \hline
         Core (C)&$0.72\pm0.07$&$-94\pm4$&$6.1\pm0.6$ &$-58.2 \pm 0.5$\\
         Northern lobe (NL) &$0.75\pm0.08$&\nodata&$18\pm2$ &\nodata \\
         Southern lobe (SL) &$1.21\pm0.08$&\nodata&$18\pm2$& \nodata\\
         Northern hotspot (NH)&$4.01\pm0.08$&\nodata&$14.7\pm0.5$&$-9.5 \pm0.2$\\
         Southern jet (SJ) &$1.85\pm0.08$&$-105\pm2$&$8.0\pm0.3$&$-170.7\pm0.2$ \\
         Northern counterjet (NCJ) &$0.31\pm0.08$&$-135\pm7$&$4.0\pm0.1$& \nodata \\
\hline
\end{tabular}
\end{table*}

\def\big{\includegraphics[trim={0 2.4cm 0 1.3cm},clip,width=17cm]{new_EVPA_1.eps}}
\def\littlea{\includegraphics[trim={5.05cm 5.32cm 7.45cm 7.87cm},clip, width=5cm,height=3.8cm, frame={1.5pt}]{new_bfield_1.eps}}
\def\littleb{\includegraphics[trim={7.7cm 8.0cm 8.35cm 6.54cm},clip, width=3.5cm,height=2.8cm, frame={1.5pt}]{PG1004+130_B.eps}}
\begin{figure*}
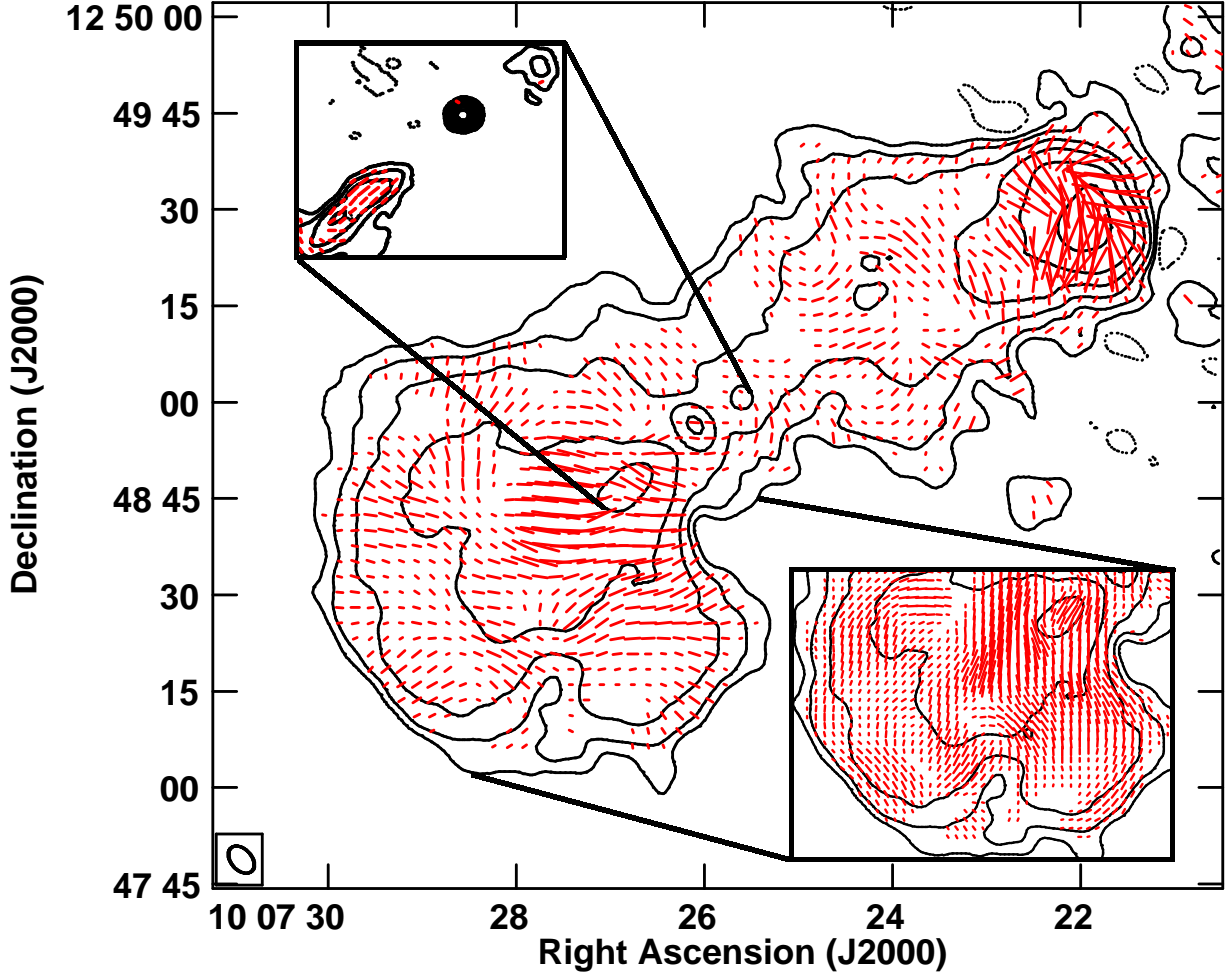

    \stackinset{l}{2.6in}{b}{.58in}{\rotatebox{-14.9}{\rule{1.75in}{1.5pt}}}{%
    \stackinset{l}{4.1in}{b}{2.1in}{\rotatebox{-9.8}{\rule{2.213in}{1.5pt}}}{%
    \stackinset{l}{1.687in}{b}{2.427in}{\rotatebox{-39}{\rule{2.07in}{1.5pt}}}{%
    \stackinset{l}{3.078in}{b}{3.03in}{\rotatebox{-62}{\rule{2.08in}{1.5pt}}}{%
    %\stackinset{l}{5.51in}{b}{1.72in}{\textcolor{blue}{\solidcirc[44]{2pt}{0.32}{0.25}}}{%
    \stackinset{r}{30pt}{b}{42pt}{\littlea}{%
    \stackinset{r}{260pt}{b}{270pt}{\littleb}{\big}}}}}}
\caption{uGMRT 694 MHz polarization image (RM-corrected) of PG1004+130 with contour levels at ($-2.83$, 2.83, 8.00, 22.63, 64.00, 181.02, 512.00) $\times$ 109 $\mu$Jy~beam$^{-1}$. The positive and negative contours are shown with solid and dashed lines, respectively. The red ticks represent the RM-corrected EVPAs. 10$\arcsec$ extent of these red vectors corresponds to a I$_P$ value of 6.25 mJy~beam$^{-1}$. The bottom-right inset displays the RM-corrected inferred B-field structure for the Southern lobe at 694 MHz with the same contours and vector sizes as in the main image. The upper-left inset shows the VLA 6 GHz B-field structure of the inner jets and the core \citep{Baghel2023}. The ellipse at the left lower corner inside the rectangle shows the synthesized beam for the image with a size of 4.86$\arcsec$ $\times$ 3.23$\arcsec$ and a position angle of 41.4$\degr$.}
    \label{fig:pol}
\end{figure*}

\begin{figure}
    \centering
    \includegraphics[width=11cm, trim=30 30 30 30]{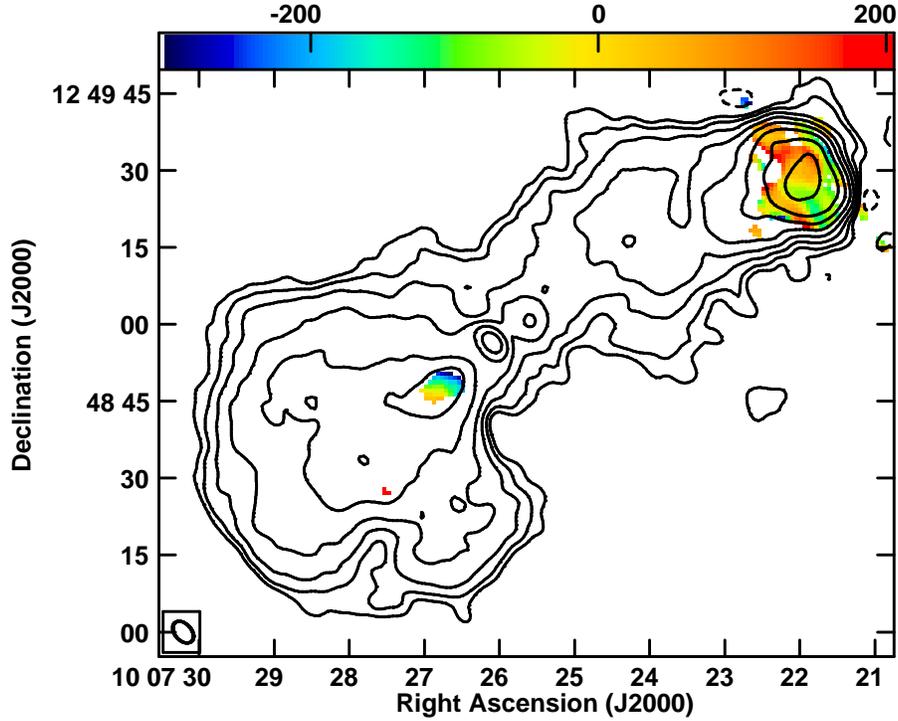}
    \caption{Rotation measure image of PG1004+130 using uGMRT 694 MHz and VLA 6 GHz (split into two sub-bands) data. The RM values are shown in colour with the colour scale extending from $-300$ to 200 rad~m$^{-2}$. The RM values are only obtained for the Southern jet and the Northern hotspot here. The contour levels are at ($-4, -2$, 4, 8, 16, 32, 64, 128, 256, 512) $\times$ 120 $\mu$Jy~beam$^{-1}$.}
    \label{fig:RM2}
\end{figure}

\subsection{Spectral Indices}\label{spectra}
Two frequency spectral index values for various components in PG1004+130 have been tabulated in Table~\ref{tab:2}. As described in Section~\ref{secspectral}, images were made at resolutions of $1.4\arcsec$ and $4\arcsec$. For compact regions like the core, Southern jet, or the Northern counterjet, we have noted spectral index values from the higher resolution image in order to eliminate the effects of the contamination by the surrounding diffuse emission. We find that at the core, the $\alpha=+0.30~\pm~0.01$ and thus the spectrum is inverted. The low resolution image shows that the Southern jet and the Northern counterjet have a steep spectrum of $\alpha=-0.91~\pm~0.02$ and $-1.4~\pm~0.1$, respectively, whereas the high resolution image shows that the $\alpha=-0.62~\pm~0.06$ for the Southern jet, while the Northern counterjet remains undetected. For calculations in the paper, we have considered the spectral index for the Northern counterjet to be the same as that of the Southern jet. The lobes show a spectral index of $\alpha=-1.2~\pm~0.1$ which is consistent with the suggestion of the lobes belonging to the previous AGN activity episode. For the Northern hotspot, the $\alpha=-0.84\pm0.06$ is somewhere between the spectral index of the jets and the lobes (see Section~\ref{sec:discussion} ahead). We note that we obtain similar values of spectral indices for the lobes ($\sim1.2\pm0.2$) over two different frequency ranges (694~MHz $-$ 1.5~GHz; 1.5~GHz $-$ 6~GHz). This indicates that the synchrotron emission does not have a spectral turnover in the frequency range, 694~MHz to 6~GHz.  

\begin{figure*}
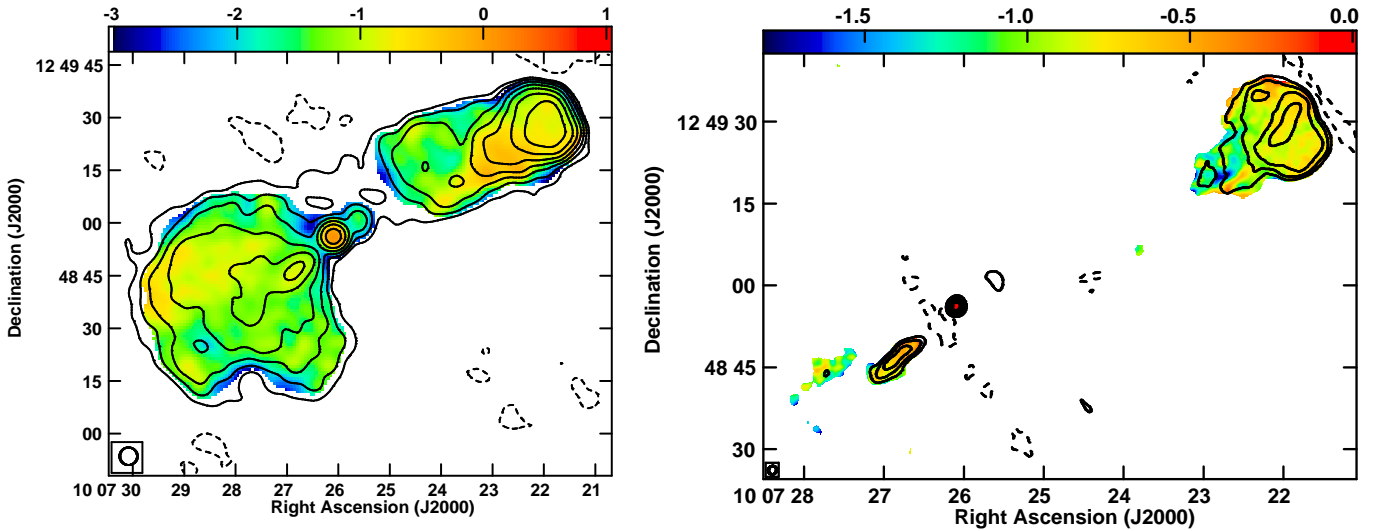

\includegraphics[width=8cm,trim=20 40 10 10]{PG1004_specind_edit.eps}
\includegraphics[width=9.8cm,trim=0 50 10 90]{1004_spix_edit.eps}
\caption{Left: The $0.694-1.5$~GHz spectral index image of PG1004+130 in colour with 1.5~GHz contour levels at $120\times (-4,4,8,16,32,64,128,256)~\mu$Jy~beam$^{-1}$. 
%The positive and negative contours are shown with solid and dashed lines, respectively. 
%The spectral index values are shown in colour with the colour scale extending from $−3.0$ to +1.0. 
The synthesized beam is of size $5\arcsec\times 5\arcsec$ at position angle of $0\degr$. Right: The $1.5-6$~GHz spectral index image of PG1004+130 in colour with 6 GHz contour levels at $120\times (-1,2,4,8,16,32,64,128)~\mu$Jy~beam$^{-1}$. 
The synthesized beam is of size $1.57\arcsec\times 1.43\arcsec$ at position angle of $-25.84\degr$.}
  \label{fig:alpha}
\end{figure*}

\subsection{Energetics}
We have estimated the equipartition magnetic field ($B_{\rm{min}}$) using the relations in \citet{O'dea1987} where the ratio of the ion energy to the electron energy ($k$), and the volume filling factor ($\phi$) was assumed to be unity. The value of the function $C_{12}$ in the equipartition calculations was taken from \citet{Pacholczyk1970}. For the estimation of the total radio luminosity \citep[see][]{O'dea1987}, we assumed the higher and lower cutoff frequencies to be 15~GHz and 100~MHz, respectively. The mean values of the radio luminosity and equipartition magnetic field have been listed in Table~\ref{tab:2}. The jet and lobe volumes were estimated assuming them to be cylindrical.

From our uGMRT image, we estimate the length of the Northern lobe to be $\sim$72$\arcsec$ ($\sim$280 kpc) with a radius of $\sim$15$\arcsec$ ($\sim$59 kpc). The Southern lobe has a length of $\sim$56$\arcsec$ ($\sim$220~kpc) and a radius of $\sim$20$\arcsec$ ($\sim$79~kpc). We note that for estimating lobe flux densities, contributions from the inner jets have been excluded. Additionally, the peak brightnesses and the `deconvolved' core and jet sizes were used for the calculations in order to avoid overestimating the flux densities due to a contamination by the surrounding diffuse lobe emission. For PG1004+130, $B_{\rm{min}}$ ranges from $\sim$ 3 to 13 $\mu$G with the maximum at the core and minimum at the lobes.

By considering the local radiation field to be CMB photons and the inverse Compton losses for CMB with a magnetic field $B_\mathrm{R} (\rm{in\,\mu G})=4(1+z)^2$ along with synchrotron radiation losses, 
the lifetime ($\tau$) of an electron radiating at a frequency $\nu$ has been provided by \citet{vanderlaan1969, O'dea1987} as $\tau (\mathrm{in\;yr})=\frac{2.6\times 10^4B^{1/2}}{(B^2+B_\mathrm{R}^2)[(1+z)\nu]^{1/2}}$. 
The average electron lifetimes for the jets and the core are smaller than that of the lobes, suggesting that the lobe electrons are much older. 
Assuming that the rotation measure resulted from a Faraday rotating screen external to the source, we can estimate the electron density in the Faraday screen using the relation, RM~$(\mathrm{in\,rad\, m^{-2}})=812\,n_e\,B_\parallel\, L$, where $B_\parallel$ is the magnetic field projected along the line of sight in $\mu$G, $n_e$ is the electron density per unit volume in cm$^{-3}$, and $L$ is the depth of Faraday rotating medium in kpc. As we do not have estimates for the magnetic fields outside the radio source and the derived $B_{\rm{min}}$ values lie in the same range as magnetic fields estimated inside galaxies \citep[e.g.,][]{Beck2004}, we have assumed $B_\parallel$ to be $B_{\rm{min}}$ \citep[see][]{Feain2009,Kharb2009,Silpa2022}. 
The electron density, here, was obtained only for those regions where we obtained RM values. This calculation gives us an electron density of around $10^{-5}-10^{-4}$ cm$^{-3}$ near the Northern hotspot region to the Southern jet (see Table~\ref{tab:2}), which is similar to that of the intergalactic medium (IGM). The $n_e$ near the core is similar to that of the interstellar medium (ISM). 

\begin{table*}
    \caption{Summary of the spectral indices and energetics for different regions of PG1004+130}
    \label{tab:2}
    \begin{center}
    \begin{tabularx}{1.15\textwidth}{lccccYYcc}
    \hhline{---------}
    \multirow{2}{*}{Region}&\multirow{2}{*}{$\alpha^{1.5\,\rm{GHz}}_{694\,\rm{MHz}}$}&\multirow{2}{*}{$\alpha^{6\,\rm{GHz}}_{1.5\,\rm{MHz}}$}&$S_C$&$L_\mathrm{rad}$ &$B_\mathrm{min}$ &$E_\mathrm{min}$&$\tau$&$n_e\,(10^{-5}$\\
    & & &(mJy)&(10$^{41}$~erg~s$^{-1}$)&(10$^{-6}$ G)&(10$^{57}$ ergs)&(10$^7$ years)&cm$^{-3}$)\\
    \hhline{---------}
    C&$+0.07\pm0.02$&$+0.30\pm0.02$&$25\pm1$&$9.9\pm0.6$&$13\pm4$&$0.4\pm0.2$&$1.5\pm0.5$&$499\pm156$\\
         %\hhline{----------}
    NL&$-1.18\pm0.09$&$-1.1\pm0.3$&$390\pm20$&$25\pm1$&$2.92\pm0.07$&$124\pm5$&$3.26\pm0.01$&$1.38\pm0.03$\\
        % \hhline{----------}
    SL&$-1.18\pm0.07$&$-1.1\pm0.2$&$910\pm45$&$57\pm3$&$3.39\pm0.07$&$231\pm8$&$3.30\pm0.01$&$1.53\pm0.04$\\
         %\hhline{----------}
    NH&$-0.79\pm0.01$&$-0.84\pm0.06$&$590\pm29$&$41\pm3$&$7.1\pm0.4$&$31\pm3$&$2.68\pm0.09$&$2.0\pm0.1$\\
        % \hhline{----------}
    SJ&$-0.91\pm0.02$&$-0.62\pm0.06$&$54\pm2$&$4.6\pm0.4$&$9\pm1$&$1.7\pm0.3$&$2.3\pm0.2$&$55\pm7$\\
       %  \hhline{----------}
    NCJ&$-1.44\pm0.07$&\nodata &$8.4\pm0.4$&$0.72\pm0.06$&$7.0\pm0.9$&$0.36\pm0.07$&$2.7\pm0.2$&\nodata\\ 
    \hhline{---------}   
    \end{tabularx}
    \end{center}

    \textit{Note:} Column 1: The different sub-structures of the QSO as observed at 694 MHz; C = core, NL = Northern lobe, SL = Southern lobe, NH = Northern hotspot, SJ = Southern jet, NCJ = Northern counterjet. Column 2: Spectral index from uGMRT 694 MHz and VLA B-array 1.5 GHz data. Column 3: Spectral index from VLA A-array 1.5 GHz and VLA B-array 6 GHz data. Column 4: Flux density at 694 MHz. For components C, SJ, and NCJ, the peak brightnesses in units of mJy~beam$^{-1}$ with a beam area of 1 steradian is tabulated. Column 5: Radio luminosity. Column 6: Magnetic field value at minimum pressure under equipartition condition. Column 7: Particle energy at minimum pressure. Column 8: Electron lifetime considering synchrotron radiation and inverse Compton losses to CMB photons. Column 9: Electron density in the Faraday-rotating medium.
\end{table*}

\subsection{Depolarization Measurements in Lobes}\label{sec:depol}
The kpc-scale uGMRT image of PG1004+130 exhibits substantial polarization throughout the extent of its lobes (Figure~\ref{fig:pol}). The absence of significant depolarizing effects even at 694~MHz offers valuable insights into the associated Faraday-rotating medium. We attempt to derive below some properties of the medium causing external and internal depolarization for PG1004+130 using the polarization information obtained at 6 GHz \citep[VLA B-array; from][]{Baghel2022} and 694 MHz (uGMRT; from the current work). 

External depolarization refers to depolarization due to Faraday rotating medium external to the source. When the synchrotron photons propagate from the source to the observer through an external medium, the FP ($p$) as a function of wavelength ($\lambda$) becomes $p(\lambda^2)=p_i\,\exp{\{-2\,(812\,n_eB_\parallel)^2\,d\,R\,\lambda^4\}}$ \citep{Burn1966}. Here, $p_i$ is the intrinsic FP of the source, `$R$' is the depth of Faraday rotating medium in kpc, `$d$' is the scale of fluctuations in the Faraday rotating medium in kpc, $n_e$ is the electron density in cm$^{-3}$, $B_\parallel$ is the line of sight magnetic field in $\mu$G, and $\lambda$ is wavelength in m \citep[see][]{Burn1966,Van1984}. The depolarizing effect due to the Galactic contribution to Faraday rotation is considered negligible here \citep{Burn1966}. 

If the external Faraday screen is comprised of clouds of hot thermal ($\sim 10^7$ K) X-ray emitting gas, we consider $n_e$ $\sim$ 10$^{-3}$ cm$^{-3}$ \citep[e.g.,][]{Luo2013, Miller2006}, $B_\parallel$ $\sim$ 3 $\mu$G (see Table \ref{tab:2}), $p(\lambda = 5$ cm) $\sim$ 30\% \citep[][for the lobes of PG1004+130]{Baghel2023}, the depth of the Faraday rotating screen (of X-ray gas) $R \sim 50\arcsec$ ($\sim$ 180 kpc) \citep{Miller2006}. From our analysis, we find $p(\lambda\approx$ 40 cm) $\sim$ 18\%. By solving the equation of external Faraday rotation for the two wavelengths, we obtain an intrinsic polarization $p_i$ of $\sim$ 30\% and the minimum size of fluctuations in the hot thermal gas $d$ as $\sim7$~pc in the lobes. If the source is surrounded by depolarizing filaments and clouds of cool ($\sim 10^4 - 10^5$ K) emission line gas, then considering $n_e$ $\approx$ 100 cm$^{-3}$ \citep{Liu2013,Harrison2014,Silpa2022}, R $\lesssim$ 0.5 kpc \citep[e.g., for 4C 26.42,][considering the nuclear line emission region]{Van1984}, we get the minimum size of the emission line clouds to be $\sim2.3\times10^{-6}$ pc, which interestingly matches the size of typical red giant stars. The half of the beam-size at 40 cm, i.e., $\sim8$~kpc here, can be considered as an upper limit to the size of the Faraday rotating clouds in both the cases, since this depolarization model assumes fluctuation scales smaller than the telescope's resolution \citep{Van1984}. 

We now consider depolarization due to internal Faraday rotation i.e., when the Faraday rotating medium (say, thermal gas) is mixed with the source radio plasma. The different amounts of Faraday rotation at different depths of the source introduce different offsets to EVPAs giving rise to front-back or internal depolarization \citep{Ioannis2015}. For this, we consider the total magnetic field ($B_{\rm{tot}}$) to be composed of a uniform magnetic field ($B_{\rm{uni}}$) and an isotropic random magnetic field ($B_{\rm{rand}}$). In this case, the complex FP as given by \citet{Burn1966} is $P(\lambda^2)=p_i\frac{1-e^{-S}}{S}$, where $S=2(812\,n_e\,B_{\rm{rand}})^2\,d\,L\,\lambda^4-1624i\,n_e\,B_{\rm{uni}}\,L\,\lambda^2$. Here, `$L$' is the linear depth of the source along the line of sight, `$d$' is the fluctuation scale of $B_{\rm{rand}}$. The intrinsic FP is given by $p_i=p_i^\prime(B_{\rm{uni}}^2/B_{\rm{tot}}^2)$, where $p_i^\prime=\frac{3-3\alpha}{5-3\alpha}$ is the theoretical intrinsic polarization for synchrotron radiation in optically thin regions when the B-fields are completely uniform \citep{Pacholczyk1970}, $\alpha$ is the spectral index, and $B_{\rm{tot}}^2=B_{\rm{uni}}^2+B_{\rm{rand}}^2$ \citep{Burn1966,Sullivan2013}. 

Using these equations and assuming negligible external depolarization, the values of $B_{\rm{uni}}$ and $B_{\rm{rand}}$ can be constrained. Assuming the intrinsic polarization from above i.e, $p_i$ as $\sim$ 30\%, and $\alpha = -1.18 \pm 0.09$ for the lobes, we obtain $p_i^\prime\approx 76\%$ and $B_{\rm{rand}}/B_{\rm{uni}}$ as $1.25\pm0.01$. Assuming an isotropic magnetic field, we obtain $B_{\rm{tot}} = \sqrt{3}B_\parallel$ \citep[see][]{Burn1966,Sullivan2013} $\sim \sqrt{3}B_{min}$. 
Therefore, we estimate $B_{\rm{rand}} = (4.26\pm0.04)\,\mu$G and $B_{\rm{uni}}=(3.42\pm0.01)\,\mu$G. The complex polarization $P(\lambda^2)$ is given by $p(\lambda^2) e^{i\chi}$, where $p(\lambda^2)$ is the FP and the $\chi$ is the polarization angle at that $\lambda$. As the lobes are large and have different $\chi$ in different regions, we consider $\chi$ in any particular region to lie between 0 to 45$\degr$ at 694 MHz (excluding rotation due to RM). We consider the depth along our line of sight through the lobes to be $L \approx 220-283$ kpc (approximated as the length of the lobes of PG1004+130), and the scale of fluctuation of random magnetic field, $d\sim 20$~kpc \citep[][for RL objects like Centaurus A]{Sullivan2013}. By solving the real and the imaginary parts of the complex polarization equation separately (using \textsc{python} package \textsc{sympy}), we estimate the electron density for the lobes as  $\approx (1.3\pm0.5)\times 10^{-5}$ cm$^{-3}$. The total mass of thermal gas present within the lobes can be estimated by $M_{tg} \sim n_e\phi m_H V$, where volume filling factor $\phi$ is assumed to be 1, total volume of the lobes $V \approx (2.17\pm0.04)\times10^{71}$ cm$^3$ (assuming the lobes as a cylinder), mass of ionised hydrogen $m_H = 8.4 \times 10^{-58}$ M$_\sun$. We obtain $M_{tg} \approx (2.4\pm0.9)\times 10^9$ M$_\sun$. We discuss the implications of these depolarization effects in Section~\ref{sec:discussion}.

\section{Discussion} \label{sec:discussion}

The uGMRT image shows the complexity of the B-field structures in PG1004+130. At the Northern hotspot, the inferred B-fields show no signs of clear compression or shock, as is typically seen in terminal hotspots \citep[e.g.,][]{Kharb2008,Perley2017b}. Rather the hotspot B-fields appear randomly oriented with respect to the jet direction and the hotspot seems relaxed, i.e., it is not compact. This is consistent with a region that is not currently being impacted by the radio jet. The inferred B-field in the inner Southern jet is parallel to the jet direction and therefore appears to be poloidal. 
\citet{Bridle1984} have suggested that the B-fields are parallel to the jet directions in FRII sources and perpendicular in FRI sources. Interestingly, the lack of a clear hotspot in the inner Southern jet could be due to the new jet propagating in the older jet plasma that results in bow-shock-like structures \citep[see][]{Clarke1992,Silpa2021}. 

In the lobes, we see `swirling' B-field structures, which may indicate turbulence inside smaller regions within the larger diffuse lobes. In the Southern lobe, we see B-field lines tracing out the edges of the lobe. Similar B-field configurations are commonly observed in FRII lobes \citep[e.g.,][]{Kharb2007,Bridle1994}. Circumferential B-field structures are also observed in some Seyfert galaxies with lobes \citep[e.g.,][]{Kharb2006,Sebastian2020}, but typically not in FRI radio galaxies where the lobe emission gradually fades away from the core \citep[e.g.,][]{Laing2011,Stanghellini1997,Feretti1999}. These circumferential field geometries indicate compression and confinement due to the external medium \citep[see][]{Laing1980}. Therefore, the uGMRT polarization data provides further evidence in support of PG1004+130 being an FRII source.

The inner jets which are detected in the Chandra X-ray observations of \citet{Miller2006} are not detected by the Hubble Space Telescope. The optical upper limit rules out a single-component synchrotron interpretation of the radio-to-X-ray emission. \citet{Miller2006} also note that the multi-wavelength characteristics of the PG1004+130 jet, including its relatively flat X-ray power law and concave spectral energy distribution, are similar to those of powerful FRII jets. On the contrary, \citet{Scott2015} suggest that the asymmetric diffuse X-ray emission that is detected along the radio lobes in PG1004+130 is indicative of a denser medium around the Southern jet/lobe, consistent with the idea that FRI jets in HYMORs are a result of greater jet-medium interaction. However, we find that the uGMRT polarization results, along with the X-ray jet emission not being single-component synchrotron emission as in FRI jets, are more consistent with the source being an FRII source. 

The visible surface brightness contrast between the inner jets and the outer lobes, where the change in surface brightness is $1.5-2.5$ times for the southern and northern sides, respectively, with respect to the ambient lobe emission, as seen in Figure~\ref{fig:total intensity}, suggests that the source PG1004+130 is a restarted source. The poloidal B-field structure in the inner Southern jet and the change of direction of the B-field vectors in the vicinity of it resembles the B-field structures observed in the radio galaxy 3C219 which has also been suggested to be a restarted AGN by \citet{Clarke1992}. Notably, no X-ray emission is detected from the northern ``hotspot'', which is consistent with the absence of an active hotspot and the suggestion of a restarted radio source. We note that \citet{Baghel2023} found the inferred B-field to be perpendicular to the jet at the base (i.e., in the VLA core) and parallel in the jet. Toroidal B-fields have been suggested to arise in AGN winds in {radio intermediate quasars} \citep{Mehdipour2019,Silpa2021} which could, in fact, be the base of the BAL wind observed in PG1004+130. Finally, we suggest that the inner Southern-jet-core-Northern-counterjet belongs to the newer AGN activity episode, whereas the Southern and Northern lobes, and the Northern hotspot belong to the older AGN activity episode. 

The newer episode seems to be FRII-like from the radio morphology as well as the polarization structure. While the previous episode appears to be like a HYMOR with diffuse emission on one side and a hotspot-like structure on the other, we suggest that the source was an FRII in the previous episode as well. There may have been a hotspot in the Southern lobe as well. However, since the hotspot decaying time-scale is short \citep[$\sim$$10^4$~years;][]{Hardcastle2001}, it may not be visible any longer. The hotspot on the northern side is still visible but is no longer compact. Considering an inclination angle of 50$\degr$ for PG1004+130 \citep{Wills1999}, a simple geometric calculation indicates that light from the Northern hotspot will take 10$^6$ years more to reach us, than from the Southern hotspot. Therefore, the Northern hotspot may have decayed/disappeared, but this information may not have reached us yet. This can explain why PG1004+130 appears to be a HYMOR even though it is actually a restarted FRII source. This result suggests that a central engine producing FRII jets in one activity episode does the same in other activity episodes. This is consistent with the studies of double-double radio galaxies in the literature \citep{Schoenmakers2000,Brocksopp2011}.

The mean spectral index of the lobes ($-1.2\pm0.1$) is consistent with the lobes (SL and NL) being from the previous AGN activity episode and the inner jets (SJ and NCJ, $\alpha=-0.62\pm0.06$) being the tracers of the present AGN activity episode. The calculated electron lifetimes for the lobes ($\approx3.3\times 10^7$ years) of the quasar are about $1.3$ times greater than of the jets ($\approx2.5\times 10^7$ years). Our calculations indicate that the age difference between the hotspot and the inner jets may be no more than $\sim7\times10^6$ years given the maximum error in the age estimation (see Table~\ref{tab:2}). The reason for the relatively small age difference might be because relativistic charged particles and fields are `leaking' from the newer jets into the older lobes. This suggestion is similar to that offered for the case of the 330~MHz radio halo seen in the famous FRI radio galaxy M87 with the VLA \citep{Owen2000,Gasperin2012}, as well as the 685 MHz `misaligned' lobe seen in the Seyfert III~Zw~2 with the uGMRT \citep{Silpa2021}, and finally the radio/X-ray lobes of the giant radio galaxy 4C23.56 as observed with the VLA/Chandra \citep{Blundell2011}. \citet{Owen2000} have suggested that in a `relativistically dominated' emitting region (i.e., dominated by fields/relativistic particles), the thermal particles penetrate into the `relativistic bubble' through semi-porous boundaries, causing a macroscopic mixing of the thermal and the relativistic plasma, while in a `thermally dominated' emitting region (i.e., dominated by thermal gas), the jet throws out its particles and fields into the ambient thermal plasma causing mixing at a microscopic level. The penetration extent of high energy ultra-relativistic particles from the jet into weakly magnetised left-over plasma can be explained by a random-walk problem as the particles flow into randomly oriented magnetic field regions \citep{Rechester1981,Blundell2011}. The thermally dominated case may be more relevant to our source. The newer episodic jet may let the energy and particles flow into the older lobes as heat or turbulence, which may, in turn, convert to magnetic energy \citep{Owen2000}. Finally, both the scenarios wherein the time difference between the jet/lobe episodes is small and charged particles are leaking from one structure into the next can coexist independently as well. 

A noteworthy similarity between PG1004+130 and M87 is that neither of the sources exhibits X-ray cavities \citep[][]{Miller2006,Owen2000}. Therefore, the X-ray emitting gas may be well mixed with the radio plasma and the strong interaction with the thermal gas may be keeping the plasma of the older episodes energised or alive. The prominent edge of the lobes may mark the boundary between the mixed relativistic and thermal plasma and the existing unperturbed X-ray gas in the IGM \citep{Owen2000}.

This mixed X-ray plasma in the lobes can give rise to internal Faraday rotation causing depolarization, whereas the hot X-ray gas clouds or cool filaments surrounding the source, outside this plasma could lead to the external depolarization. The intrinsic FP values as obtained for the lobes of PG1004+130, is similar to the FP obtained at 6~GHz by \citet[][see Section~\ref{sec:depol} above]{Baghel2023}, implying that the effects due to external Faraday rotation are negligible in the case of higher frequencies like 6 GHz. This might also suggest that the sizes of some of the Faraday rotating clouds or filaments along our line of sight to the source are greater than the beam-size at 6 GHz \citep[see][]{Van1984}, i.e, of the order of a few kpc. The drop in the FP values at 694 MHz with the uGMRT observations (resolution $\sim4\arcsec$) compared to the 6 GHz VLA observations of \citet{Baghel2023} (resolution $\sim1.4\arcsec$) can be explained by beam and frequency-dependent (external) depolarization. 

Since we detect contiguous polarized regions in the lobes and jets even at 694 MHz with the uGMRT, it may indicate that the randomization of the B-field vectors due to propagation effects (internal or external), is low. We find the mass of the thermal gas mixed with the radio plasma causing internal depolarization to be $\sim(2.4\pm0.9)\times 10^9$ M$_\sun$. Additionally, this mass estimate falls in the range of thermal gas mass ($10^9$ to $10^{10}$~M$_\sun$) obtained for other RL sources in the literature \citep[e.g.,][]{Kirkpatrick2009, Salome2011,Sullivan2013}.

\section{Summary and Conclusions}
We have presented the first polarization-sensitive images of the hybrid morphology BAL quasar PG1004+130 at 694~MHz on kpc-scales with the uGMRT. We have detected extensive diffuse and polarized radio emission in the lobes and jets of this source. The main findings of this study are the following.

\begin{enumerate}
 \item The uGMRT 694 MHz total intensity image shows a clear distinction between the inner jet-core-counterjet structure and the Southern and Northern lobes. Therefore, we suggest PG1004+130 to be a restarted AGN with two activity episodes, with the inner jet structures belonging to the current AGN activity episode and the larger outer lobes to the previous activity episode. 

\item The spectral index image shows that the core has an inverted spectrum ($\alpha=+0.30\pm0.02$), the inner jets have a steep spectrum ($\alpha=-0.62\pm0.06$), while the outer lobes have much steeper spectra ($\alpha\approx-1.2\pm0.1$). These results are consistent with PG1004+130 being a restarted AGN with two episodes of activity. The electron lifetimes of the lobes and the jets show the (maximum) time difference between the two episodes is $\sim1.2\times 10^7$ years. The relatively small age difference compared to the ages themselves may also indicate that charges and fields are `leaking' from the new jets into the older lobes, keeping them largely active.

\item The FP varies between 4 to 18\% across the source. The FP in the core is $6.1~\pm~0.6$\%, whereas the Northern hotspot shows a FP value of $14.7~\pm~0.5$\%. The inferred B-fields in the Northern hotspot appear to be randomly oriented with respect to the jet direction and are not reminiscent of compressed B-fields that are typically observed in terminal hotspots. This is suggestive of a relaxed B-field configuration consistent with a hotspot that is not currently being impacted by a jet. The inferred B-fields in the inner southern jet are parallel to the jet direction, as observed in FRII jets. The B-field in the lobes shows swirling features hinting at turbulence and mixing of plasma. Moreover, the B-fields align with the Southern lobe edge, similar to what is usually observed in FRII lobes. 

\item  Using uGMRT 694~MHz and VLA 6~GHz polarization data, we estimate the value of intrinsic polarization for the lobes to be $\sim 30\%$. The external Faraday screen may be made up of the clumpy and hot X-ray gas clouds of sizes $7-8000$ parsec, or filaments of line-emitting cool gas of sizes $2.3\times10^{-6}-8000$ parsec, where half the beam-size at 694 MHz is considered as the upper limit to the sizes of the clouds. The internal depolarization is not strong enough to completely depolarize the lobes of the source. The mass of thermal gas which could be mixed with the radio plasma causing internal depolarization is around $(2.4\pm 0.9)\times10^9$ M$_\sun$.

\item Based on the inferred B-fields in the Southern lobe and the short lifetimes of hotspots ($\sim10^4$~yr) compared to the significant light travel time ($\sim$$10^{6}$~yr) from the northern end to the southern end of this large radio source, we conclude that PG1004+130 was an FRII source in its previous AGN activity episode. Based on the inferred poloidal B-fields in the inner Southern jet and the Chandra X-ray data suggesting that it is not from single-component synchrotron emission, which is typically seen in FRI jets, we deduce that the current AGN activity episode is also FRII. The lack of a clear hotspot in the inner southern jet is due to the new jet propagating in the older jet plasma resulting in bow-shock-like structures. Therefore, we conclude that an FRII source continues to remain an FRII source in multiple activity episodes. 
\end{enumerate}
Radio polarization is therefore an invaluable tool for understanding the dichotomy of AGN jets and their environment. Moreover, this study demonstrates the capability of uGMRT polarimetric observations in inferring the kpc-scale B-field structures in complex AGNs like PG1004+130.

\section*{Acknowledgements}
We thank the referee for their suggestions that have improved this manuscript considerably. SG, PK, JB, and SS acknowledge the support of the Department of Atomic Energy, Government of India, under the project 12-R\&D-TFR-5.02-0700. We thank the staff of the GMRT that made these observations possible. GMRT is run by the National Centre for Radio Astrophysics of the Tata Institute of Fundamental Research. This work has made use of observations carried out with the Karl G. Jansky Very Large Array. 
The National Radio Astronomy Observatory is a facility of the National Science Foundation operated under cooperative agreement by Associated Universities, Inc. This research has made use of the NASA/IPAC Extragalactic Database (NED), which is operated by the Jet Propulsion Laboratory, California Institute of Technology, under contract with the National Aeronautics and Space Administration.

%%%%%%%%%%%%%%%%%%%%%%%%%%%%%%%%%%%%%%%%%%%%%%%%%%
\section*{Data Availability}
The data underlying this article will be shared on reasonable request to the corresponding author. The uGMRT data underlying this article can be obtained from the GMRT Archive (http://naps.ncra.tifr.res.in/goa/data/search) using the proposal ID: 37\_042. The VLA data underlying
this article can be obtained from the NRAO Data Archive
(https://science.nrao.edu/facilities/vla/archive/index) using the project
ID: 20A-182, AK219, and WARD.

\bibliography{sample631}{}

\begin{thebibliography}{}
\expandafter\ifx\csname natexlab\endcsname\relax\def\natexlab#1{#1}\fi
\providecommand{\url}[1]{\href{#1}{#1}}
\providecommand{\dodoi}[1]{doi:~\href{http://doi.org/#1}{\nolinkurl{#1}}}
\providecommand{\doeprint}[1]{\href{http://ascl.net/#1}{\nolinkurl{http://ascl.net/#1}}}
\providecommand{\doarXiv}[1]{\href{https://arxiv.org/abs/#1}{\nolinkurl{https://arxiv.org/abs/#1}}}

\bibitem[{{Baghel} {et~al.}(2023){Baghel}, {Kharb}, {Silpa}, {Ho}, \& {Harrison}}]{Baghel2023}
{Baghel}, J., {Kharb}, P., {Silpa}, {Ho}, L.~C., \& {Harrison}, C.~M. 2023, \mnras, 519, 2773, \dodoi{10.1093/mnras/stac3691}

\bibitem[{{Baghel} {et~al.}(2022){Baghel}, {Silpa}, {Kharb}, {Sebastian}, \& {Shastri}}]{Baghel2022}
{Baghel}, J., {Silpa}, S., {Kharb}, P., {Sebastian}, B., \& {Shastri}, P. 2022, Journal of Astrophysics and Astronomy, 43, 88, \dodoi{10.1007/s12036-022-09879-8}

\bibitem[{{Beck}(2004)}]{Beck2004}
{Beck}, R. 2004, in Astrophysics and Space Science Library, Vol. 315, How Does the Galaxy Work?, ed. E.~J. {Alfaro}, E.~{P{\'e}rez}, \& J.~{Franco}, 277, \dodoi{10.1007/1-4020-2620-X_57}

\bibitem[{{Blandford} {et~al.}(2019){Blandford}, {Meier}, \& {Readhead}}]{Blandford2019}
{Blandford}, R., {Meier}, D., \& {Readhead}, A. 2019, \araa, 57, 467, \dodoi{10.1146/annurev-astro-081817-051948}

\bibitem[{{Blandford} \& {Payne}(1982)}]{Blandford1982}
{Blandford}, R.~D., \& {Payne}, D.~G. 1982, \mnras, 199, 883, \dodoi{10.1093/mnras/199.4.883}

\bibitem[{{Blandford} \& {Znajek}(1977)}]{Blandford1977}
{Blandford}, R.~D., \& {Znajek}, R.~L. 1977, \mnras, 179, 433, \dodoi{10.1093/mnras/179.3.433}

\bibitem[{{Blundell} \& {Fabian}(2011)}]{Blundell2011}
{Blundell}, K.~M., \& {Fabian}, A.~C. 2011, \mnras, 412, 705, \dodoi{10.1111/j.1365-2966.2010.17608.x}

\bibitem[{{Bridle}(1982)}]{Bridle1982}
{Bridle}, A.~H. 1982, in Extragalactic Radio Sources, ed. D.~S. {Heeschen} \& C.~M. {Wade}, Vol.~97, 121--128

\bibitem[{{Bridle}(1984)}]{Bridle1984}
{Bridle}, A.~H. 1984, \aj, 89, 979, \dodoi{10.1086/113593}

\bibitem[{{Bridle} {et~al.}(1994){Bridle}, {Hough}, {Lonsdale}, {Burns}, \& {Laing}}]{Bridle1994}
{Bridle}, A.~H., {Hough}, D.~H., {Lonsdale}, C.~J., {Burns}, J.~O., \& {Laing}, R.~A. 1994, \aj, 108, 766, \dodoi{10.1086/117112}

\bibitem[{{Brocksopp} {et~al.}(2011){Brocksopp}, {Kaiser}, {Schoenmakers}, \& {de Bruyn}}]{Brocksopp2011}
{Brocksopp}, C., {Kaiser}, C.~R., {Schoenmakers}, A.~P., \& {de Bruyn}, A.~G. 2011, \mnras, 410, 484, \dodoi{10.1111/j.1365-2966.2010.17456.x}

\bibitem[{{Burn}(1966)}]{Burn1966}
{Burn}, B.~J. 1966, \mnras, 133, 67, \dodoi{10.1093/mnras/133.1.67}

\bibitem[{{Clarke} {et~al.}(1992){Clarke}, {Bridle}, {Burns}, {Perley}, \& {Norman}}]{Clarke1992}
{Clarke}, D.~A., {Bridle}, A.~H., {Burns}, J.~O., {Perley}, R.~A., \& {Norman}, M.~L. 1992, \apj, 385, 173, \dodoi{10.1086/170925}

\bibitem[{Contopoulos {et~al.}(2015)Contopoulos, Gabuzda, \& Kylafis}]{Ioannis2015}
Contopoulos, I., Gabuzda, D., \& Kylafis, N., eds. 2015, Astrophysics and Space Science Library, Vol. 414, {The Formation and Disruption of Black Hole Jets}, ed. I.~Contopoulos, D.~Gabuzda, \& N.~Kylafis (Springer), \dodoi{10.1007/978-3-319-10356-3}

\bibitem[{{Croston} {et~al.}(2018){Croston}, {Ineson}, \& {Hardcastle}}]{Croston2018}
{Croston}, J.~H., {Ineson}, J., \& {Hardcastle}, M.~J. 2018, \mnras, 476, 1614, \dodoi{10.1093/mnras/sty274}

\bibitem[{{de Gasperin} {et~al.}(2012){de Gasperin}, {Orr{\'u}}, {Murgia}, {Merloni}, {Falcke}, {Beck}, {Beswick}, {B{\^\i}rzan}, {Bonafede}, {Br{\"u}ggen}, {Brunetti}, {Chy{\.z}y}, {Conway}, {Croston}, {En{\ss}lin}, {Ferrari}, {Heald}, {Heidenreich}, {Jackson}, {Macario}, {McKean}, {Miley}, {Morganti}, {Offringa}, {Pizzo}, {Rafferty}, {R{\"o}ttgering}, {Shulevski}, {Steinmetz}, {Tasse}, {van der Tol}, {van Driel}, {van Weeren}, {van Zwieten}, {Alexov}, {Anderson}, {Asgekar}, {Avruch}, {Bell}, {Bell}, {Bentum}, {Bernardi}, {Best}, {Breitling}, {Broderick}, {Butcher}, {Ciardi}, {Dettmar}, {Eisloeffel}, {Frieswijk}, {Gankema}, {Garrett}, {Gerbers}, {Griessmeier}, {Gunst}, {Hassall}, {Hessels}, {Hoeft}, {Horneffer}, {Karastergiou}, {K{\"o}hler}, {Koopman}, {Kuniyoshi}, {Kuper}, {Maat}, {Mann}, {Mevius}, {Mulcahy}, {Munk}, {Nijboer}, {Noordam}, {Paas}, {Pandey}, {Pandey}, {Polatidis}, {Reich}, {Schoenmakers}, {Sluman}, {Smirnov}, {Sobey}, {Stappers}, {Swinbank}, {Tagger}, {Tang}, {van Bemmel}, {van Cappellen},
  {van Duin}, {van Haarlem}, {van Leeuwen}, {Vermeulen}, {Vocks}, {White}, {Wise}, {Wucknitz}, \& {Zarka}}]{Gasperin2012}
{de Gasperin}, F., {Orr{\'u}}, E., {Murgia}, M., {et~al.} 2012, \aap, 547, A56, \dodoi{10.1051/0004-6361/201220209}

\bibitem[{{Elvis}(2000)}]{Elvis2000}
{Elvis}, M. 2000, \apj, 545, 63, \dodoi{10.1086/317778}

\bibitem[{{Fanaroff} \& {Riley}(1974)}]{Fanaroff1974}
{Fanaroff}, B.~L., \& {Riley}, J.~M. 1974, \mnras, 167, 31P, \dodoi{10.1093/mnras/167.1.31P}

\bibitem[{{Farnes} {et~al.}(2014){Farnes}, {Green}, \& {Kantharia}}]{Farnes2014}
{Farnes}, J.~S., {Green}, D.~A., \& {Kantharia}, N.~G. 2014, \mnras, 437, 3236, \dodoi{10.1093/mnras/stt2118}

\bibitem[{{Feain} {et~al.}(2009){Feain}, {Ekers}, {Murphy}, {Gaensler}, {Macquart}, {Norris}, {Cornwell}, {Johnston-Hollitt}, {Ott}, \& {Middelberg}}]{Feain2009}
{Feain}, I.~J., {Ekers}, R.~D., {Murphy}, T., {et~al.} 2009, \apj, 707, 114, \dodoi{10.1088/0004-637X/707/1/114}

\bibitem[{{Feretti} {et~al.}(1999){Feretti}, {Perley}, {Giovannini}, \& {Andernach}}]{Feretti1999}
{Feretti}, L., {Perley}, R., {Giovannini}, G., \& {Andernach}, H. 1999, \memsai, 70, 129

\bibitem[{{Fomalont}(1981)}]{Fomalont1981}
{Fomalont}, E.~B. 1981, in Origin of Cosmic Rays, ed. G.~{Setti}, G.~{Spada}, \& A.~W. {Wolfendale}, Vol.~94, 111--125

\bibitem[{{Gawro{\'n}ski} {et~al.}(2006){Gawro{\'n}ski}, {Marecki}, {Kunert-Bajraszewska}, \& {Kus}}]{Gawronski2006}
{Gawro{\'n}ski}, M.~P., {Marecki}, A., {Kunert-Bajraszewska}, M., \& {Kus}, A.~J. 2006, \aap, 447, 63, \dodoi{10.1051/0004-6361:20053996}

\bibitem[{{Goodrich} \& {Miller}(1995)}]{Goodrich1995}
{Goodrich}, R.~W., \& {Miller}, J.~S. 1995, \apjl, 448, L73, \dodoi{10.1086/309600}

\bibitem[{{Gopal-Krishna} \& {Wiita}(2000)}]{Gopal-Krishna2000}
{Gopal-Krishna}, \& {Wiita}, P.~J. 2000, \aap, 363, 507.
\newblock \doarXiv{astro-ph/0009441}

\bibitem[{{Gopal-Krishna} {et~al.}(2023){Gopal-Krishna}, {Joshi}, \& {Patra}}]{Gopal-Krishna2023}
{Gopal-Krishna}, Wiita, P.~J., {Joshi}, R., \& {Patra}, D. 2023, Journal of Astrophysics and Astronomy, 44, 44, \dodoi{10.1007/s12036-023-09937-9}

\bibitem[{{Hall} {et~al.}(2013){Hall}, {Brandt}, {Petitjean}, {P{\^a}ris}, {Filiz Ak}, {Shen}, {Gibson}, {Aubourg}, {Anderson}, {Schneider}, {Bizyaev}, {Brinkmann}, {Malanushenko}, {Malanushenko}, {Myers}, {Oravetz}, {Ross}, {Shelden}, {Simmons}, {Streblyanska}, {Weaver}, \& {York}}]{Hall2013}
{Hall}, P.~B., {Brandt}, W.~N., {Petitjean}, P., {et~al.} 2013, \mnras, 434, 222, \dodoi{10.1093/mnras/stt1012}

\bibitem[{{Hardcastle} \& {Looney}(2001)}]{Hardcastle2001}
{Hardcastle}, M.~J., \& {Looney}, L.~W. 2001, \mnras, 320, 355, \dodoi{10.1046/j.1365-8711.2001.03963.x}

\bibitem[{{Harrison} {et~al.}(2014){Harrison}, {Alexander}, {Mullaney}, \& {Swinbank}}]{Harrison2014}
{Harrison}, C.~M., {Alexander}, D.~M., {Mullaney}, J.~R., \& {Swinbank}, A.~M. 2014, \mnras, 441, 3306, \dodoi{10.1093/mnras/stu515}

\bibitem[{{Harwood} {et~al.}(2020){Harwood}, {Vernstrom}, \& {Stroe}}]{Harwood2020}
{Harwood}, J.~J., {Vernstrom}, T., \& {Stroe}, A. 2020, \mnras, 491, 803, \dodoi{10.1093/mnras/stz3069}

\bibitem[{{Kale} {et~al.}(2019){Kale}, {Shende}, \& {Parekh}}]{Kale2019}
{Kale}, R., {Shende}, K.~M., \& {Parekh}, V. 2019, \mnras, 486, L80, \dodoi{10.1093/mnrasl/slz061}

\bibitem[{{Kapi{\'n}ska} {et~al.}(2017){Kapi{\'n}ska}, {Terentev}, {Wong}, {Shabala}, {Andernach}, {Rudnick}, {Storer}, {Banfield}, {Willett}, {de Gasperin}, {Lintott}, {L{\'o}pez-S{\'a}nchez}, {Middelberg}, {Norris}, {Schawinski}, {Seymour}, \& {Simmons}}]{Kapinska2017}
{Kapi{\'n}ska}, A.~D., {Terentev}, I., {Wong}, O.~I., {et~al.} 2017, \aj, 154, 253, \dodoi{10.3847/1538-3881/aa90b7}

\bibitem[{{Kellermann} {et~al.}(1989){Kellermann}, {Sramek}, {Schmidt}, {Shaffer}, \& {Green}}]{Kellermann1989}
{Kellermann}, K.~I., {Sramek}, R., {Schmidt}, M., {Shaffer}, D.~B., \& {Green}, R. 1989, \aj, 98, 1195

\bibitem[{{Kharb} {et~al.}(2009){Kharb}, {Gabuzda}, {O'Dea}, {Shastri}, \& {Baum}}]{Kharb2009}
{Kharb}, P., {Gabuzda}, D.~C., {O'Dea}, C.~P., {Shastri}, P., \& {Baum}, S.~A. 2009, \apj, 694, 1485, \dodoi{10.1088/0004-637X/694/2/1485}

\bibitem[{{Kharb} {et~al.}(2010){Kharb}, {Lister}, \& {Cooper}}]{Kharb2010}
{Kharb}, P., {Lister}, M.~L., \& {Cooper}, N.~J. 2010, \apj, 710, 764, \dodoi{10.1088/0004-637X/710/1/764}

\bibitem[{{Kharb} {et~al.}(2006){Kharb}, {O'Dea}, {Baum}, {Colbert}, \& {Xu}}]{Kharb2006}
{Kharb}, P., {O'Dea}, C.~P., {Baum}, S.~A., {Colbert}, E.~J.~M., \& {Xu}, C. 2006, \apj, 652, 177, \dodoi{10.1086/507945}

\bibitem[{{Kharb} {et~al.}(2008{\natexlab{a}}){Kharb}, {O'Dea}, {Baum}, {Daly}, {Mory}, {Donahue}, \& {Guerra}}]{Kharb2008}
{Kharb}, P., {O'Dea}, C.~P., {Baum}, S.~A., {et~al.} 2008{\natexlab{a}}, \apjs, 174, 74, \dodoi{10.1086/520840}

\bibitem[{{Kharb} {et~al.}(2008{\natexlab{b}}){Kharb}, {O'Dea}, {Baum}, {Daly}, {Mory}, {Donahue}, \& {Guerra}}]{Kharb2007}
---. 2008{\natexlab{b}}, \apjs, 174, 74, \dodoi{10.1086/520840}

\bibitem[{{Kirkpatrick} {et~al.}(2009){Kirkpatrick}, {McNamara}, {Rafferty}, {Nulsen}, {B{\^\i}rzan}, {Kazemzadeh}, {Wise}, {Gitti}, \& {Cavagnolo}}]{Kirkpatrick2009}
{Kirkpatrick}, C.~C., {McNamara}, B.~R., {Rafferty}, D.~A., {et~al.} 2009, \apj, 697, 867, \dodoi{10.1088/0004-637X/697/1/867}

\bibitem[{{Kumari} \& {Pal}(2022)}]{Sobha2022}
{Kumari}, S., \& {Pal}, S. 2022, \mnras, 514, 4290, \dodoi{10.1093/mnras/stac1215}

\bibitem[{{Kunert-Bajraszewska} {et~al.}(2013){Kunert-Bajraszewska}, {Katarzy{\'n}ski}, {Janiuk}, \& {Ceg{\l}owski}}]{Kunert-Bajraszewska2013}
{Kunert-Bajraszewska}, M., {Katarzy{\'n}ski}, K., {Janiuk}, A., \& {Ceg{\l}owski}, M. 2013, in Feeding Compact Objects: Accretion on All Scales, ed. C.~M. {Zhang}, T.~{Belloni}, M.~{M{\'e}ndez}, \& S.~N. {Zhang}, Vol. 290, 243--244, \dodoi{10.1017/S1743921312019825}

\bibitem[{{Laing}(1980)}]{Laing1980}
{Laing}, R.~A. 1980, in Bulletin of the American Astronomical Society, Vol.~12, 823

\bibitem[{{Laing} \& {Bridle}(2002)}]{Laing2002}
{Laing}, R.~A., \& {Bridle}, A.~H. 2002, \mnras, 336, 328, \dodoi{10.1046/j.1365-8711.2002.05756.x}

\bibitem[{{Laing} {et~al.}(2011){Laing}, {Guidetti}, {Bridle}, {Parma}, \& {Bondi}}]{Laing2011}
{Laing}, R.~A., {Guidetti}, D., {Bridle}, A.~H., {Parma}, P., \& {Bondi}, M. 2011, \mnras, 417, 2789, \dodoi{10.1111/j.1365-2966.2011.19436.x}

\bibitem[{{Liu} {et~al.}(2013){Liu}, {Zakamska}, {Greene}, {Nesvadba}, \& {Liu}}]{Liu2013}
{Liu}, G., {Zakamska}, N.~L., {Greene}, J.~E., {Nesvadba}, N. P.~H., \& {Liu}, X. 2013, \mnras, 436, 2576, \dodoi{10.1093/mnras/stt1755}

\bibitem[{{Luo} {et~al.}(2013){Luo}, {Brandt}, {Alexander}, {Harrison}, {Stern}, {Bauer}, {Boggs}, {Christensen}, {Comastri}, {Craig}, {Fabian}, {Farrah}, {Fiore}, {Fuerst}, {Grefenstette}, {Hailey}, {Hickox}, {Madsen}, {Matt}, {Ogle}, {Risaliti}, {Saez}, {Teng}, {Walton}, \& {Zhang}}]{Luo2013}
{Luo}, B., {Brandt}, W.~N., {Alexander}, D.~M., {et~al.} 2013, \apj, 772, 153, \dodoi{10.1088/0004-637X/772/2/153}

\bibitem[{{Mehdipour} \& {Costantini}(2019)}]{Mehdipour2019}
{Mehdipour}, M., \& {Costantini}, E. 2019, \aap, 625, A25, \dodoi{10.1051/0004-6361/201935205}

\bibitem[{{Miller} {et~al.}(2006){Miller}, {Brandt}, {Gallagher}, {Laor}, {Wills}, {Garmire}, \& {Schneider}}]{Miller2006}
{Miller}, B.~P., {Brandt}, W.~N., {Gallagher}, S.~C., {et~al.} 2006, \apj, 652, 163, \dodoi{10.1086/507509}

\bibitem[{{Mingo} {et~al.}(2019){Mingo}, {Croston}, {Hardcastle}, {Best}, {Duncan}, {Morganti}, {Rottgering}, {Sabater}, {Shimwell}, {Williams}, {Brienza}, {Gurkan}, {Mahatma}, {Morabito}, {Prandoni}, {Bondi}, {Ineson}, \& {Mooney}}]{Mingo2019}
{Mingo}, B., {Croston}, J.~H., {Hardcastle}, M.~J., {et~al.} 2019, \mnras, 488, 2701, \dodoi{10.1093/mnras/stz1901}

\bibitem[{{O'Brien}(1998)}]{O'Brien1998}
{O'Brien}, P. 1998, The Observatory, 118, 337

\bibitem[{{O'Dea} \& {Owen}(1987)}]{O'dea1987}
{O'Dea}, C.~P., \& {Owen}, F.~N. 1987, \apj, 316, 95, \dodoi{10.1086/165182}

\bibitem[{{O'Sullivan} {et~al.}(2013){O'Sullivan}, {Feain}, {McClure-Griffiths}, {Ekers}, {Carretti}, {Robishaw}, {Mao}, {Gaensler}, {Bland-Hawthorn}, \& {Stawarz}}]{Sullivan2013}
{O'Sullivan}, S.~P., {Feain}, I.~J., {McClure-Griffiths}, N.~M., {et~al.} 2013, \apj, 764, 162, \dodoi{10.1088/0004-637X/764/2/162}

\bibitem[{{Owen} {et~al.}(2000){Owen}, {Eilek}, \& {Kassim}}]{Owen2000}
{Owen}, F.~N., {Eilek}, J.~A., \& {Kassim}, N.~E. 2000, \apj, 543, 611, \dodoi{10.1086/317151}

\bibitem[{{Pacholczyk}(1970)}]{Pacholczyk1970}
{Pacholczyk}, A.~G. 1970, {Radio astrophysics} (%San Francisco : W.H. Freeman, cop. 1970)

\bibitem[{{Perley} \& {Butler}(2017)}]{Perley2017}
{Perley}, R.~A., \& {Butler}, B.~J. 2017, \apjs, 230, 7, \dodoi{10.3847/1538-4365/aa6df9}

\bibitem[{{Perley} \& {Meisenheimer}(2017)}]{Perley2017b}
{Perley}, R.~A., \& {Meisenheimer}, K. 2017, \aap, 601, A35, \dodoi{10.1051/0004-6361/201629704}

\bibitem[{{Rau} \& {Cornwell}(2011)}]{Rau2011}
{Rau}, U., \& {Cornwell}, T.~J. 2011, \aap, 532, A71, \dodoi{10.1051/0004-6361/201117104}

\bibitem[{Rechester {et~al.}(1981)Rechester, Rosenbluth, \& White}]{Rechester1981}
Rechester, A.~B., Rosenbluth, M.~N., \& White, R.~B. 1981, Phys. Rev. A, 23, 2664, \dodoi{10.1103/PhysRevA.23.2664}

\bibitem[{{Rees}(1984)}]{Rees1984}
{Rees}, M.~J. 1984, \araa, 22, 471, \dodoi{10.1146/annurev.aa.22.090184.002351}

\bibitem[{{Reichard} {et~al.}(2004){Reichard}, {Richards}, {Hall}, \& {Schneider}}]{Reichard2004}
{Reichard}, T., {Richards}, G., {Hall}, P., \& {Schneider}, D. 2004, in Astronomical Society of the Pacific Conference Series, Vol. 311, AGN Physics with the Sloan Digital Sky Survey, ed. G.~T. {Richards} \& P.~B. {Hall}, 219

\bibitem[{{Salom{\'e}} {et~al.}(2011){Salom{\'e}}, {Combes}, {Revaz}, {Downes}, {Edge}, \& {Fabian}}]{Salome2011}
{Salom{\'e}}, P., {Combes}, F., {Revaz}, Y., {et~al.} 2011, \aap, 531, A85, \dodoi{10.1051/0004-6361/200811333}

\bibitem[{{Schoenmakers} {et~al.}(2000){Schoenmakers}, {de Bruyn}, {R{\"o}ttgering}, {van der Laan}, \& {Kaiser}}]{Schoenmakers2000}
{Schoenmakers}, A.~P., {de Bruyn}, A.~G., {R{\"o}ttgering}, H.~J.~A., {van der Laan}, H., \& {Kaiser}, C.~R. 2000, \mnras, 315, 371, \dodoi{10.1046/j.1365-8711.2000.03430.x}

\bibitem[{{Scott} {et~al.}(2015){Scott}, {Brandt}, {Miller}, {Luo}, \& {Gallagher}}]{Scott2015}
{Scott}, A.~E., {Brandt}, W.~N., {Miller}, B.~P., {Luo}, B., \& {Gallagher}, S.~C. 2015, \apj, 806, 210, \dodoi{10.1088/0004-637X/806/2/210}

\bibitem[{{Sebastian} {et~al.}(2020){Sebastian}, {Kharb}, {O'Dea}, {Gallimore}, \& {Baum}}]{Sebastian2020}
{Sebastian}, B., {Kharb}, P., {O'Dea}, C.~P., {Gallimore}, J.~F., \& {Baum}, S.~A. 2020, \mnras, 499, 334, \dodoi{10.1093/mnras/staa2473}

\bibitem[{{Shangguan} {et~al.}(2018){Shangguan}, {Ho}, \& {Xie}}]{Shangguan2018}
{Shangguan}, J., {Ho}, L.~C., \& {Xie}, Y. 2018, \apj, 854, 158, \dodoi{10.3847/1538-4357/aaa9be}

\bibitem[{{Silpa} {et~al.}(2022){Silpa}, {Kharb}, {Harrison}, {Girdhar}, {Mukherjee}, {Mainieri}, \& {Jarvis}}]{Silpa2022}
{Silpa}, S., {Kharb}, P., {Harrison}, C.~M., {et~al.} 2022, \mnras, 513, 4208, \dodoi{10.1093/mnras/stac1044}

\bibitem[{{Silpa} {et~al.}(2021){Silpa}, {Kharb}, {Harrison}, {Ho}, {Jarvis}, {Ishwara-Chandra}, \& {Sebastian}}]{Silpa2021}
---. 2021, \mnras, 507, 991, \dodoi{10.1093/mnras/stab1870}

\bibitem[{{Simard-Normandin} {et~al.}(1981){Simard-Normandin}, {Kronberg}, \& {Button}}]{Simard1981}
{Simard-Normandin}, M., {Kronberg}, P.~P., \& {Button}, S. 1981, \apjs, 45, 97, \dodoi{10.1086/190709}

\bibitem[{{Stanghellini} {et~al.}(1997){Stanghellini}, {Dallacasa}, {Bondi}, \& {della Ceca}}]{Stanghellini1997}
{Stanghellini}, C., {Dallacasa}, D., {Bondi}, M., \& {della Ceca}, R. 1997, \aap, 325, 911

\bibitem[{{Stocke} {et~al.}(1992){Stocke}, {Morris}, {Weymann}, \& {Foltz}}]{Stocke1992}
{Stocke}, J.~T., {Morris}, S.~L., {Weymann}, R.~J., \& {Foltz}, C.~B. 1992, \apj, 396, 487, \dodoi{10.1086/171735}

\bibitem[{{Stroe} {et~al.}(2022){Stroe}, {Catlett}, {Harwood}, {Vernstrom}, \& {Mingo}}]{Stroe2022}
{Stroe}, A., {Catlett}, V., {Harwood}, J.~J., {Vernstrom}, T., \& {Mingo}, B. 2022, arXiv e-prints, arXiv:2210.09315.
\newblock \doarXiv{2210.09315}

\bibitem[{{THE CASA TEAM} {et~al.}(2022){THE CASA TEAM}, {Bean}, {Bhatnagar}, {Castro}, {Donovan Meyer}, {Emonts}, {Garcia}, {Garwood}, {Golap}, {Gonzalez Villalba}, {Harris}, {Hayashi}, {Hoskins}, {Hsieh}, {Jagannathan}, {Kawasaki}, {Keimpema}, {Kettenis}, {Lopez}, {Marvil}, {Masters}, {McNichols}, {Mehringer}, {Miel}, {Moellenbrock}, {Montesino}, {Nakazato}, {Ott}, {Petry}, {Pokorny}, {Raba}, {Rau}, {Schiebel}, {Schweighart}, {Sekhar}, {Shimada}, {Small}, {Steeb}, {Sugimoto}, {Suoranta}, {Tsutsumi}, {van Bemmel}, {Verkouter}, {Wells}, {Xiong}, {Szomoru}, {Griffith}, {Glendenning}, \& {Kern}}]{CASA2022}
{THE CASA TEAM}, {Bean}, B., {Bhatnagar}, S., {et~al.} 2022, arXiv e-prints, arXiv:2210.02276.
\newblock \doarXiv{2210.02276}

\bibitem[{{van Breugel} {et~al.}(1984){van Breugel}, {Heckman}, \& {Miley}}]{Van1984}
{van Breugel}, W., {Heckman}, T., \& {Miley}, G. 1984, \apj, 276, 79, \dodoi{10.1086/161594}

\bibitem[{{van der Laan} \& {Perola}(1969)}]{vanderlaan1969}
{van der Laan}, H., \& {Perola}, G.~C. 1969, \aap, 3, 468

\bibitem[{{van Moorsel} {et~al.}(1996){van Moorsel}, {Kemball}, \& {Greisen}}]{vanMoorsel1996}
{van Moorsel}, G., {Kemball}, A., \& {Greisen}, E. 1996, in Astronomical Society of the Pacific Conference Series, Vol. 101, Astronomical Data Analysis Software and Systems V, ed. G.~H. {Jacoby} \& J.~{Barnes}, 37

\bibitem[{{Vantyghem} {et~al.}(2014){Vantyghem}, {McNamara}, {Russell}, {Main}, {Nulsen}, {Wise}, {Hoekstra}, \& {Gitti}}]{Vantyghem2014}
{Vantyghem}, A.~N., {McNamara}, B.~R., {Russell}, H.~R., {et~al.} 2014, \mnras, 442, 3192, \dodoi{10.1093/mnras/stu1030}

\bibitem[{{Wang} {et~al.}(2023){Wang}, {An}, {Cheng}, {Ho}, {Kellermann}, {Baan}, {Yang}, \& {Zhang}}]{Wang2023}
{Wang}, A., {An}, T., {Cheng}, X., {et~al.} 2023, \mnras, 518, 39, \dodoi{10.1093/mnras/stac3091}

\bibitem[{{Willis} {et~al.}(1981){Willis}, {Strom}, {Bridle}, \& {Fomalont}}]{Willis1981}
{Willis}, A.~G., {Strom}, R.~G., {Bridle}, A.~H., \& {Fomalont}, E.~B. 1981, \aap, 95, 250

\bibitem[{{Wills} {et~al.}(1999){Wills}, {Brandt}, \& {Laor}}]{Wills1999}
{Wills}, B.~J., {Brandt}, W.~N., \& {Laor}, A. 1999, \apjl, 520, L91, \dodoi{10.1086/312165}

\bibitem[{{Wright} \& {Otrupcek}(1990)}]{Parkescatalog1990}
{Wright}, A., \& {Otrupcek}, R. 1990, PKS Catalog (1990, 0

\end{thebibliography}
\bibliographystyle{aasjournal}

\end{document}